\documentclass[final,5p,times,authoryear,twocolumn]{elsarticle}
\usepackage{amssymb}
\usepackage{graphicx}
\usepackage{graphics}
\usepackage[centerlast,sc,small]{subfigure}
\usepackage[dvipsnames]{xcolor}
\usepackage{slashbox}
\usepackage{amsmath}
\usepackage{color}
\usepackage{colortbl}
\usepackage{appendix}
\usepackage{mathptmx} 

\usepackage{txfonts}
\usepackage[authoryear]{natbib}

\journal{Planetary and Space Science}

\begin{document}

\begin{frontmatter}



\title{Dynamics of asteroids and near-Earth objects from Gaia Astrometry}


\author[1]{D. Bancelin}
\author[1]{D. Hestroffer}
\author[1]{W. Thuillot}

\address[1]{              IMCCE, Paris Observatory, CNRS, UPMC\\ 
              77, Av. Denfert-Rochereau 75014 Paris France \\}

\begin{abstract}
Gaia is an astrometric mission that will be launched in spring 2013. There are
many scientific outcomes from this mission and as far as our Solar 
System is concerned, the satellite will be able to map thousands of main belt
asteroids (MBAs) and near-Earth objects (NEOs) down to magnitude $\le 20$. The
high precision astrometry ($0.3-5$ mas of accuracy) will allow orbital
improvement, mass determination, and a better accuracy in the prediction and
ephemerides of potentially hazardous asteroids (PHAs).\\ 
We give in this paper some simulation tests to analyse the impact of Gaia data
on known asteroids's orbit, and their value for the analysis of NEOs through the
example of asteroid 
(99942) Apophis. We then present the need for a follow-up network for newly
discovered asteroids by Gaia, insisting on the synergy of ground and space data
for the orbital improvement.

\end{abstract}


\end{frontmatter}

% \linenumbers

\section{Introduction}
\label{S:Intro}

Science of asteroids and comets, from near-Earth objects (NEOs) to
trans-Neptunian objects (TNOs), and small bodies of the Solar System at large is
fundamental to understand the formation and evolution of the Solar System
starting from the proto-Sun and the planetary embryos. Having little geological
evolution and being atmosphere free, their pristine character makes them good
tracers of the constitution of the primordial Solar System. Being numerous and
spread over a wide range of heliocentric distances they act also as good
constraints for planetary formation scenario and the Solar System dynamical
evolution at large. Last, knowledge of the process within our Solar System is
useful if not mandatory to understand formations and evolution of other
planetary system than our own Solar System.

While some objects can be considered as small world on their own, such as targets of space probes, the vast majority will be considered
through general groups and classes. Some asteroids are planet crossers or evolving in the vicinity of Earth's orbit. Among the latter, a
small fraction of potentially hazardous asteroids (PHAs) can show particuliar threat of collision with the Earth while others have no
incidence at all. Near-Earth objects are also of interest to understand the physics process as non-gravitational forces (in particular the
Yarkovsky effect) and fundamental physics with local tests of General Relativity.

%--------------------------------------------------------------------------------------------

\section{Gaia detection and observations of asteroids}
\label{S:detec}
Gaia will observe a large number of asteroids, however with some specificity and
limits. The limiting magnitude is modest when compared to present and future
ground-based surveys aimed at making a census of small bodies\footnote{The
objectif reclaimed to the NASA by the US congress is to catalogue 90\% of NEOs
larger than 140\,m.}. On another hand Gaia will enable observations with a
single instrument of the entire celestial sphere and also at low solar
elongation, making a difference between space-based observations -- such as
AsteroidFinder \citep{mottola10_cosp} and NEOSSat \citep{hildebrand04}-- and
typical ground-based observations and surveys. As seen in \cite{mignard07}, the
Gaia satellite will have a peculiar scanning law enabling a full coverage of the
entire sky over 6\,months, which coverage is repeated over the 5\,years mission
providing stellar parallaxes and proper motions. Besides, only objects detected
and confirmed in the front CCDs forming the sky mapper will be subsequently
observed through the main astrometric field-of-view. This ensures that no cosmic
rays are treated as scientific sources and enables to download to ground only
small windows around a scientific source and not all the pixels of the large CCD
mosaic. Nevertheless the detection algorithm is so that extended sources, when too wide, are not detected by the on-board
algorithm. As shown in Fig. 1, there is no clear detection limit,  solar system objects in the size range 0.7-0.9 arcsec will not be systematically
detected, while objects larger than 0.9 arcsec will not be observed.

\begin{figure}[htbp]
\centerline{
	\includegraphics[width=\hsize]{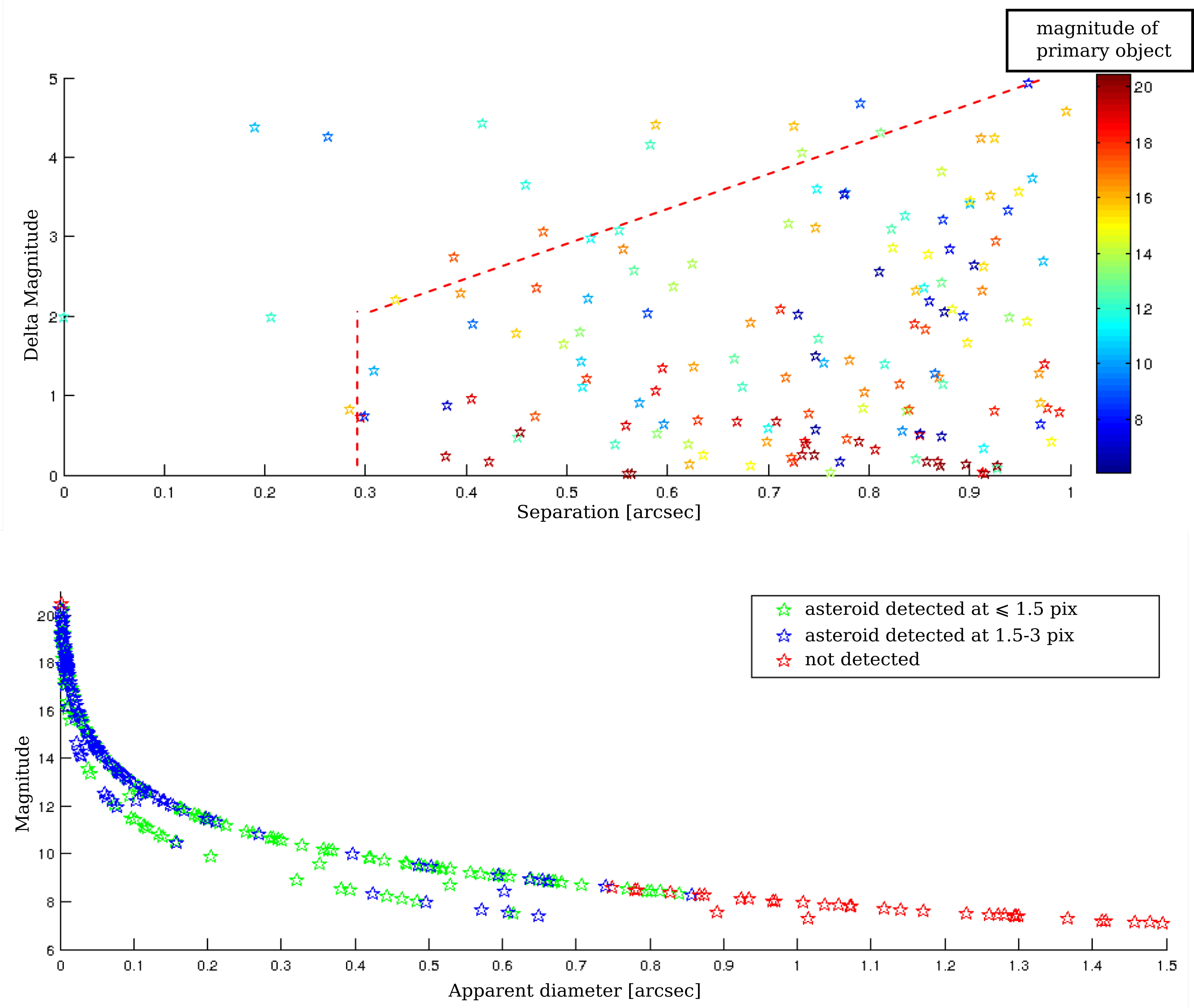}
}
\caption{Detection limits, in the sky mapper, for binary objects (top) and large asteroids (bottom). Top panel: the detection is given as a function
of the separation of the pair (irrespective of its position angle) and the magnitude difference between the secondary and the primary; the colour
code
indicates the magnitude of the primary. The detection in the binned sky mapper CCDs stops at a separation of less than approximately 0.3\,arcsec
(corresponding to $\approx 2.5$ binned pixel). Bottom panel: the detection is given as function of the apparent diameter of the object. The
corresponding apparent magnitude is derived for a given albedo and three different heliocentric distances. Objects larger than 0.7\,arcsec will not
be
systematically detected; when detected, their predicted position can show an offset from the true one by several pixels. }
\label{F:detection}
\end{figure}

The sequence of observation of any object hence depends on this scanning law, the on-board detection, and the limiting magnitude. Starting with the
{\tt astorb} database \citep{bowell94} of orbital elements, one can compute dates of rendez-vous of asteroids crossing the Gaia FOV with the CU4 Solar
System Simulator. Simulations in the focal plane of images making use of the GIBIS tool \citep{luri11} will enable to set the detection of large
asteroids and planetary satellites. Making use of the GIBIS tool, Fig.~\ref{F:detection} shows some detection limits for binary objects and large
asteroids. These are, in the case of binary systems, the detection in the SM CCD that are binned (2x2) and hence of lower resolution. In such case
each component will be treated individually with an associated patch and windowing for observation in the subsequent CCDs. While not detected at the
SM level, binary systems can still be observed in the AF field, with higher, but basically one dimensional patches resolution (personal
communication).
Concerning large asteroids, it appears that even Ceres and some planetary satellites will be basically detected and observed.

%%------------------------------------------------------
%%------------------------------------------------------

Statistics on observations of asteroids have been reported in \cite{mignard07,hestro10_lnp}. On the average there are 60 transits (or
observations) per object over the mission duration. Fast moving objects will not be observed correctly through the whole astrometric field of view
because the windowing scheme is adapted to the relative motion of a star (personal communication). Objects like fast NEOs will however be observed in
good
conditions in the first and middle CCDs (which has a larger associated window).

%--------------------------------------------------------------------------------------------

\section{Dynamic of asteroids}
\label{S: detec}

\begin{figure}[htbp]
\centerline{
	\includegraphics[width=\hsize]{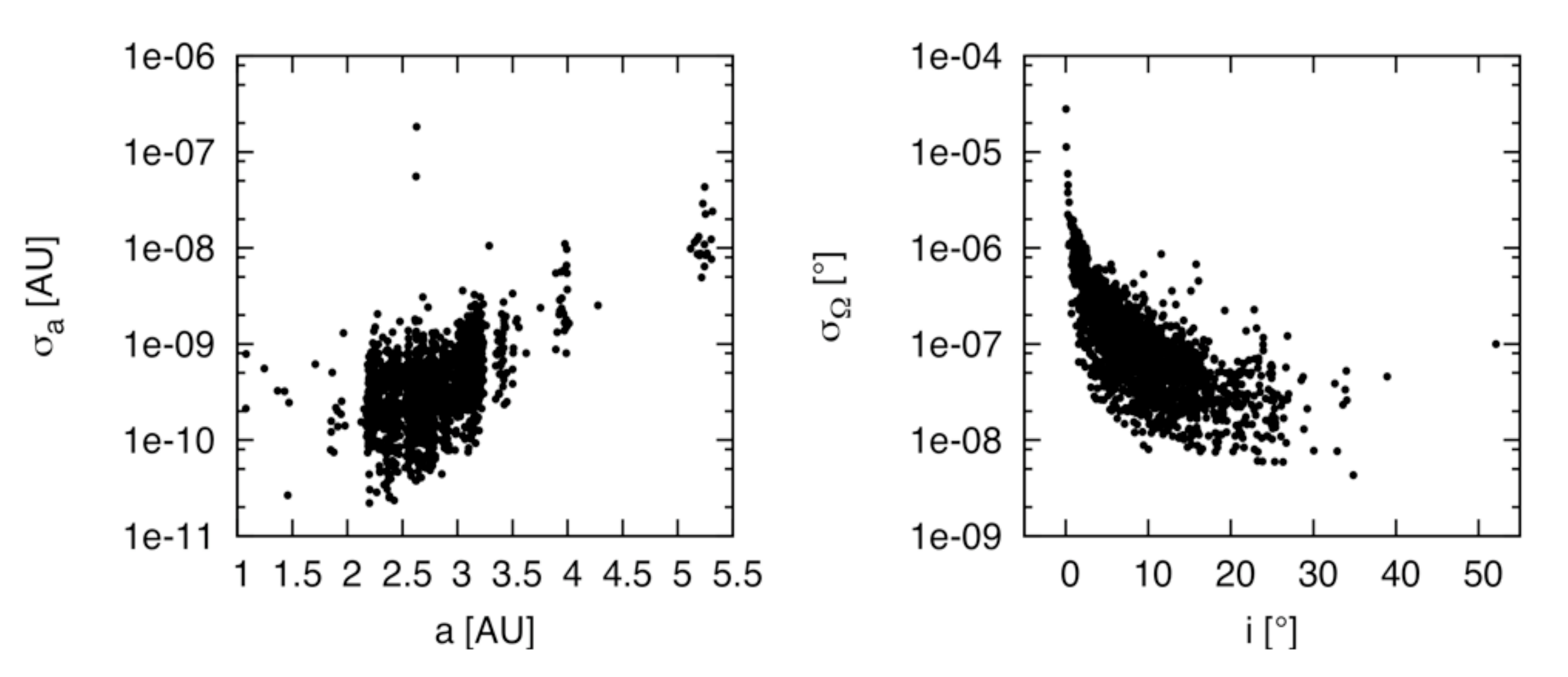}
}
\caption{Orbit improvement in semi-major axis and orbital plane orientation. The improvement is given as the
precision on the correction to the state vector or orbital elements from a linear least squares fit from Gaia observations alone. For a small
percentage
of
objects the number of observations and/or their distribution will be too small to derive a complete orbit (rank deficiency in the linear least
squares
inversion), but for the vast majority the astrometric precision of the order of 0.3--5\,mas will enable orbit improvement by factor $\approx10-50$.}
\label{F:orbitPrec}
\end{figure}

%DH\\
%\textit{
%    Determination\\
%    Estimation des masses\\
%    Tests de la RG}
Gaia will provide astrometry of asteroids and comets with unprecedented accuracy. Being a space-mission designed optimally for doing astrometry it has
some obvious advantages. Gaia will in particular enable both local astrometry from relative positions and refined calibration, and global astrometry
with absolute positions. Compared to classical ground-based observations, there are---among other---no limitation between northern and southern
hemisphere, no atmospheric refraction or turbulent effects, reduced zonal errors, and positions directly in the Gaia sphere of reference and the
optical ICRF. Such astrometry will yield improved orbital elements for almost all objects observed (see Fig.~\ref{F:orbitPrec}), together with
detection of small effects and determination of dynamical parameters. 
In particular, one will be able to derive masses of asteroids (from close encounter and binary objects) and to perform local tests of general
relativity (GR). 
We do not consider here preliminary orbit determination for newly detected objects that will be treated in
Sect.~\ref{S:network}, neither dynamics of planetary satellites that will not be treated within DPAC with Gaia data alone.

%%------------------------------------------------------
%%------------------------------------------------------

\begin{figure}[htbp]
\centerline{
	\includegraphics[width=\hsize]{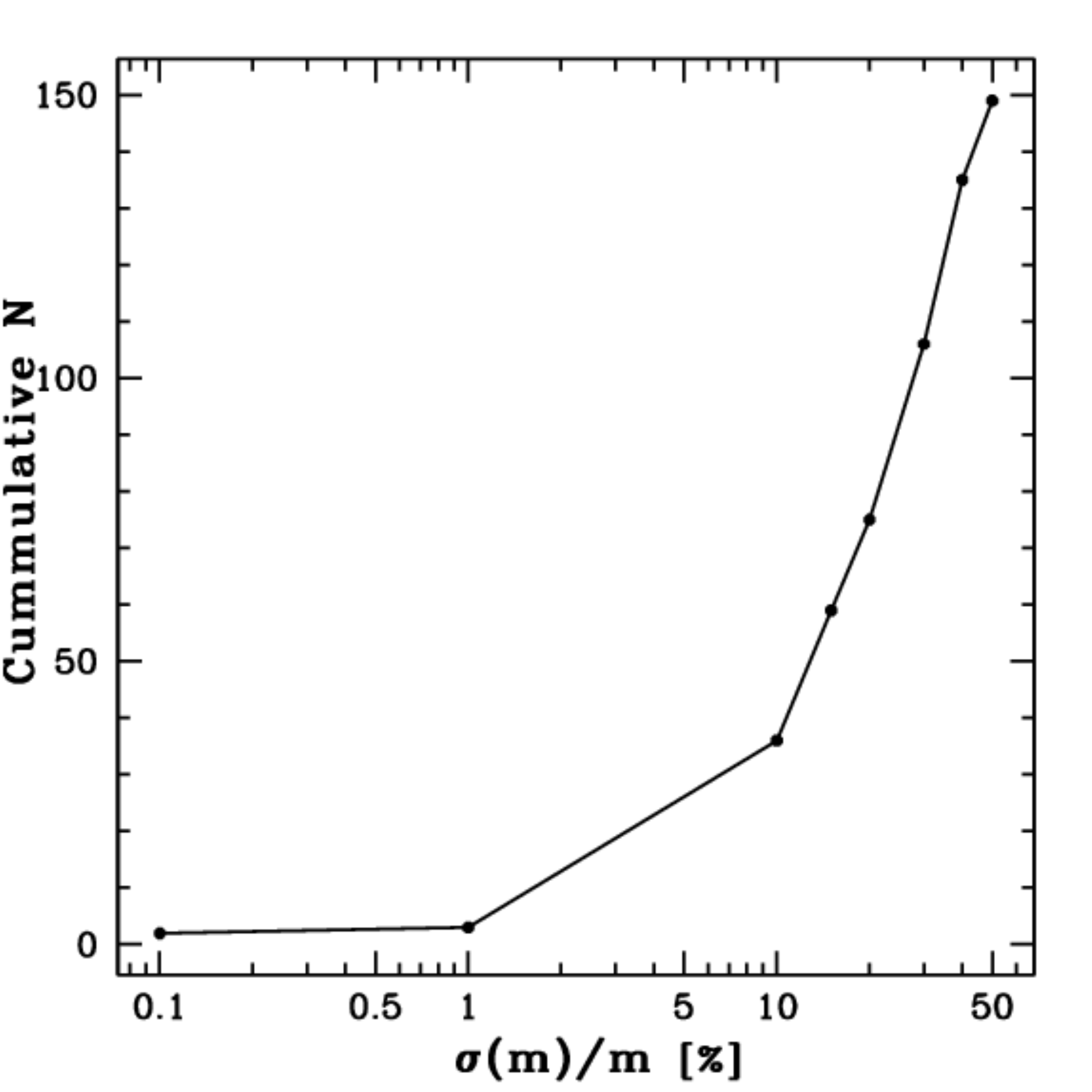}
}
\caption{Mass determination from close encounters. Cumulative distribution as a function of the relative precision reached \citep{mouret07}.}
\label{F:mass}
\end{figure}

The mass of an asteroid can be measured during a close encounter from the trajectory's deflection caused on a perturbed smaller body \citep{hilton02}.
The situation is improved and less subject to systematic errors when several perturbers per perturber asteroid are involved. In the Gaia data
processing scheme a preselected list of perturber asteroids has been done based on computation of close encounters during the mission
\citep{mouret07}. Simulation of a global inversion of the problem involving 43\,500 perturbed targets and 600 massive asteroids (in 78\,800 close
approaches) has shown that 150 asteroids (i.e. $\approx25\%$) could have their mass derived to better than 50\% \citep{mouret07_phd}, see
Fig.~\ref{F:mass}. There are 36 asteroids with their mass determined to better than 10\% (including Vesta and Ceres that are presently observed by the
Dawn mission, and some binary asteroids) distributed in several taxonomic classes. This number is slightly increased when complementing the Gaia
observations by ground-based data for those close encounters that happen just before or after the mission \citep{hudkova08}. Good knowledge of their
volume will be mandatory to derive reliable estimates of their bulk density and further indication of their porosity (personal communication).
With a pixel size of 0.06\,arcsec, observations of some resolved binary systems will also be possible with Gaia---though one dimensional---including
the Pluton/Charon system for which the relative positions acquired over 5 years will provide substantial refinement of the knowledge of the system's
mass \citep{beauvalet_pisa}. On the other hand, Gaia can also detect in principle astrometric binaries, this has to be investigated further.
%%------------------------------------------------------
%%------------------------------------------------------

The refined orbits of asteroids will also provide valuable inputs for local tests of General Relativity, basically derivation of the PPN parameter
$\beta$ \citep{will10_iaus} from monitoring the precession of perihelion of eccentric NEOs (i.e. large eccentricity $e$, small semi-major axis $a$)
together with the derivation of the solar quadrupole $J_{2}$. Additionally all asteroids will contribute to a test of a possible time variation of the
gravitational constant $dG/dt$ and a possible residual rotation $d{\mathbf W}/dt$ between the kinematic reference frame materialised by the QSO and
the dynamical reference frame materialised by the motion of the asteroids. It has been shown in \cite{hestro10_iaus} that---due to the good $(e,a)$
plane coverage, good monitoring of both perihelion $\omega$, node $\Omega$ and inclination $I$, and the large number of test particles involved---the
parameters $\beta$ and $J_2$ will be derived individually to a precision of $\approx 10^{-4}$ and $\approx 10^{-8}$, respectively. Such precision is
similar to what is obtained from other techniques, yet independently and directly without assumptions on the Sun interior or the Nordvedt parameter.
Combination of Gaia astrometry of NEOs to radar data \citep{margot10_iaus} can in principle bring a higher time leverage for measuring this secular
effect, this has to be investigated further.

It is worth to mention that the Gaia data alone from direct observation astrometry of Solar System objects can yield scientific outputs as shown
above, but it can also complement ground-based data over long time span. Last, the Gaia catalogue of stars will provide the astrometry of tomorrow
including re-reduction or debiasing of ancient CCD observations, better prediction of stellar occultation, and dense catalogue for small fields
reduction without severe zonal errors.

%--------------------------------------------------------------------------------------------

\section{Observations of PHAs}
\label{S:PHAs}

\begin{figure}[htbp]
\centerline{
	\includegraphics[width=\hsize]{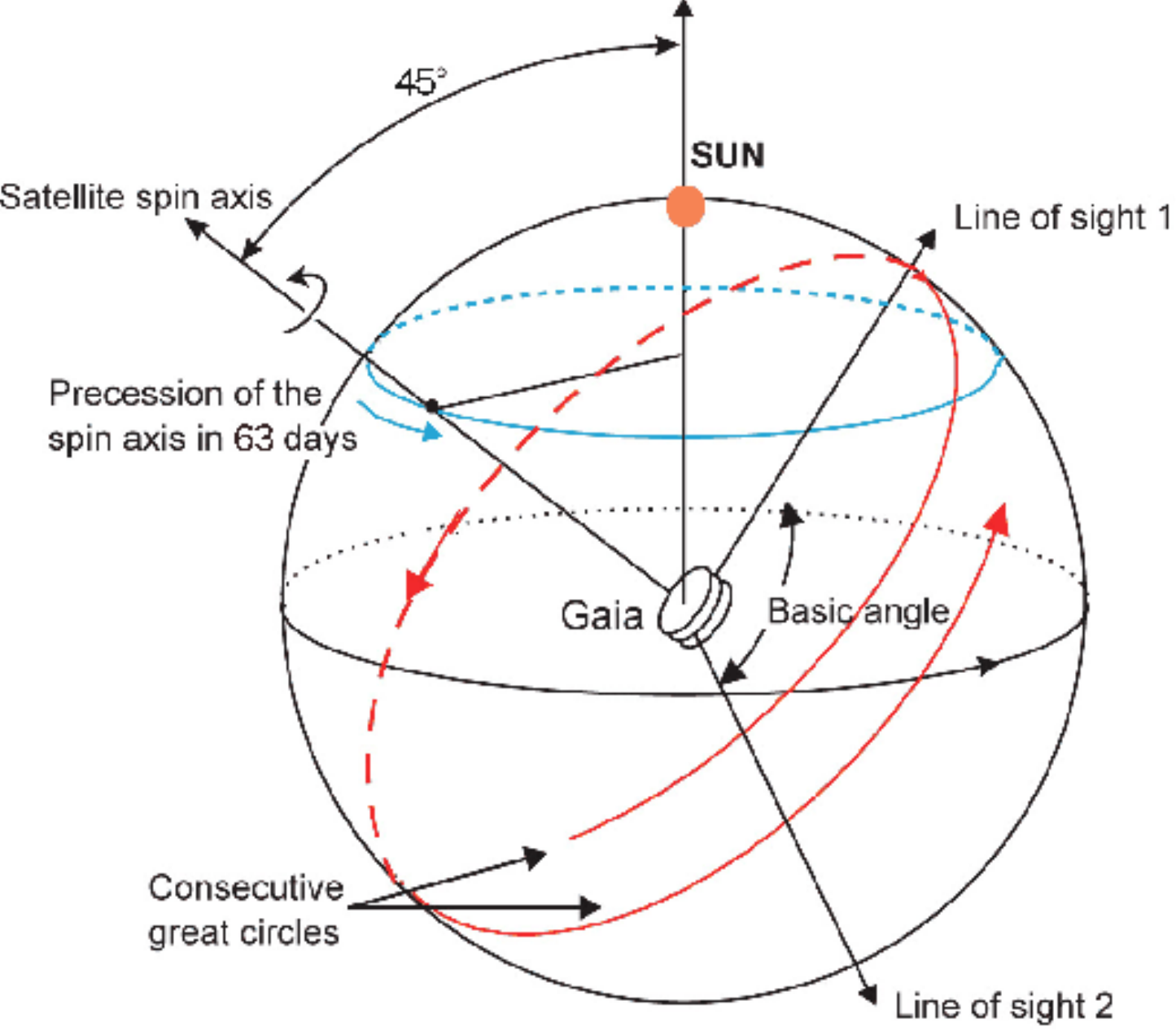}
}
\caption{Nominal scanning law of Gaia (Source: ESA). Six parameters determine this scanning law: the basic-angle (angle between the two lines of
sight), the inertial spin rate (angular speed of the spacecraft), the solar-aspect angle (angle between the Sun direction and the satellite spin
axis), the precession period (rotation of the spacecraft around the Sun-Earth direction, the inertial spin phase and the initial precession angle.}
\label{F:nominal}
\end{figure}

During the 5 years mission, Gaia will continously scan the sky with a specific strategy as shown in Fig.~\ref{F:nominal}: Objects will be observed
from two lines of sight separated with a constant basic angle. Some constants already fixed determine the nominal scanning law of Gaia:
The inertial spin rate ($1^{\circ}/$min) that describes the rotation of the spacecraft around an axis perpendicular to those of the two fields of
view, the solar-aspect angle ($45^{\circ}$) that is the angle between the Sun and the spacecraft spin axis, the precession period ($63.12$ days) 
which is the precession of the spin axis around the Sun-Earth direction. Two other constants are still free parameters: the initial spin phase which
has an influence on the observation's dates and the initial precession angle which has an influence on the number of observations for a given target.
Those parameters will be fixed at the start of the nominal science operations. These latter are constrained by scientific outcome (e.g. possibility of
performing test of fundamental physics) together with operational requirements (downlink to Earth windows).\\

% --------------------------------------------------------------------------------------------------------------
%---------------------------------------------------------------------------------------------------------------

\begin{figure}[htbp!]
\centerline{
	\includegraphics[width=\hsize]{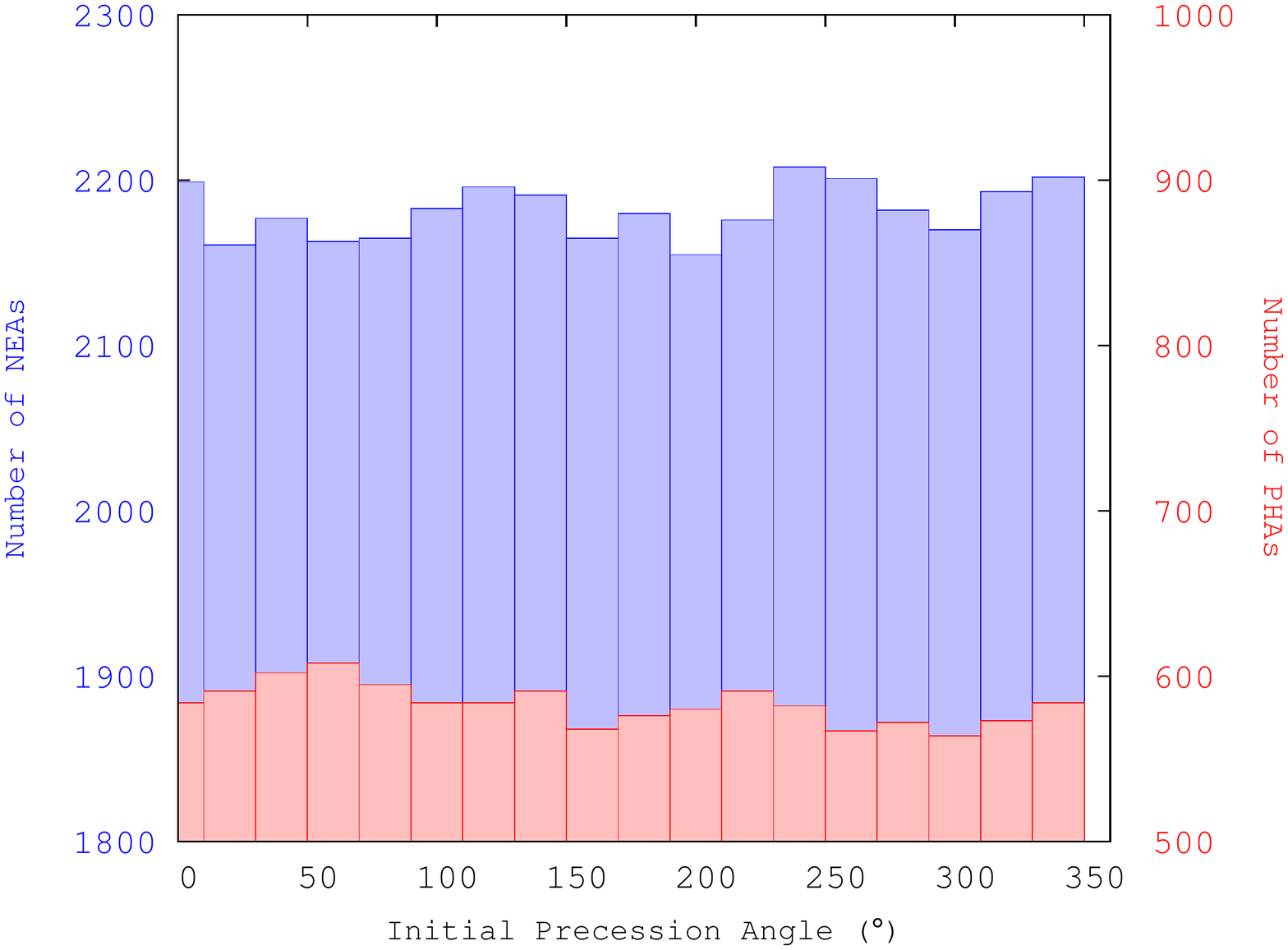}
}
\caption{Number of NEOs and PHAs that would be observed by Gaia with respect to the initial precession angle. Only $30\%$ of the NEOs population
could
be
observed by Gaia. Amoung the most hazardous population, the PHAs, Gaia would observe only $1/4$ of them.}
\label{F:histo}
\end{figure}

\begin{figure}[htbp!]
\centerline{
	\includegraphics[width=\hsize]{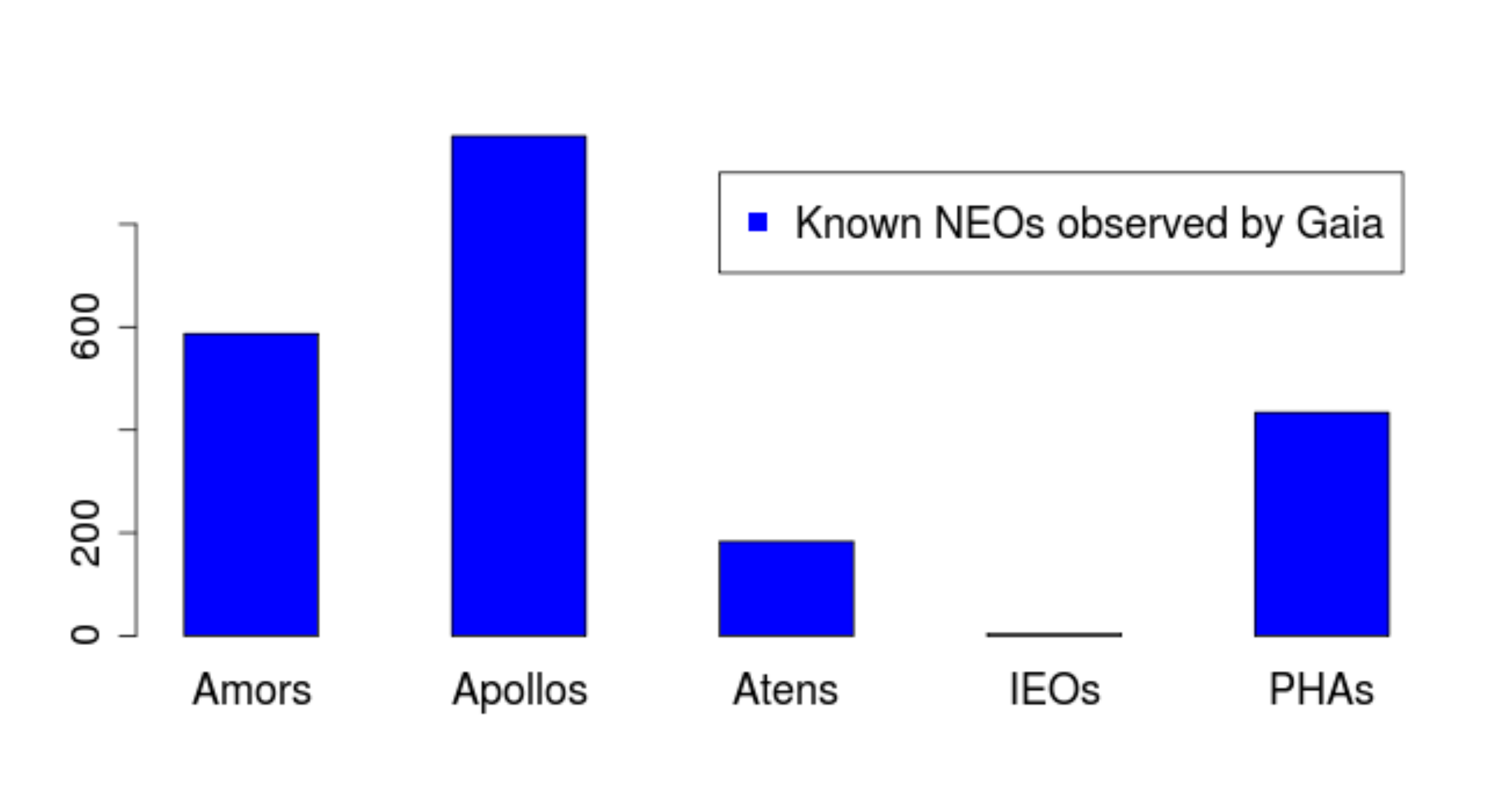}
}
\caption{Statistic of observations for the near-Earth objects with respect to their dynamical family. The weak variation of the number of observed
objects with respect to the initial precession angle as seen in Fig. \ref{F:histo}, justifies to consider a mean value of the population possibly
observed by Gaia.}
\label{F:histo_nea}
\end{figure}

Different sequences of observations of NEOs will be possible according to the initial value of the initial precession angle. Figure
\ref{F:histo} is an histogram
showing the number of NEOs and PHAs that would be observed by the satellite (an object is considered to be observed at the first detection).
We can first see that the number of NEOs that could be observed is weak
compared to the population of knows NEOs ($\sim 30\%$). Besides, the number of objects observed do not vary greatly regarding the initial
precession
angle. As a matter of fact, the mean value and standard deviation for each distribution is $2180\pm16$ NEOs observed by Gaia and $585\pm12$
for PHAs. So we can just give the mean value of objects that would be observed, regarding their dynamical family as shown on Fig. \ref{F:histo_nea}. 

% --------------------------------------------------------------------------------------------------------------
%---------------------------------------------------------------------------------------------------------------

% --------------------------------------------------------------------------------------------------------------
%---------------------------------------------------------------------------------------------------------------

\begin{figure}[htbp!]
\centerline{
	\includegraphics[width=\hsize]{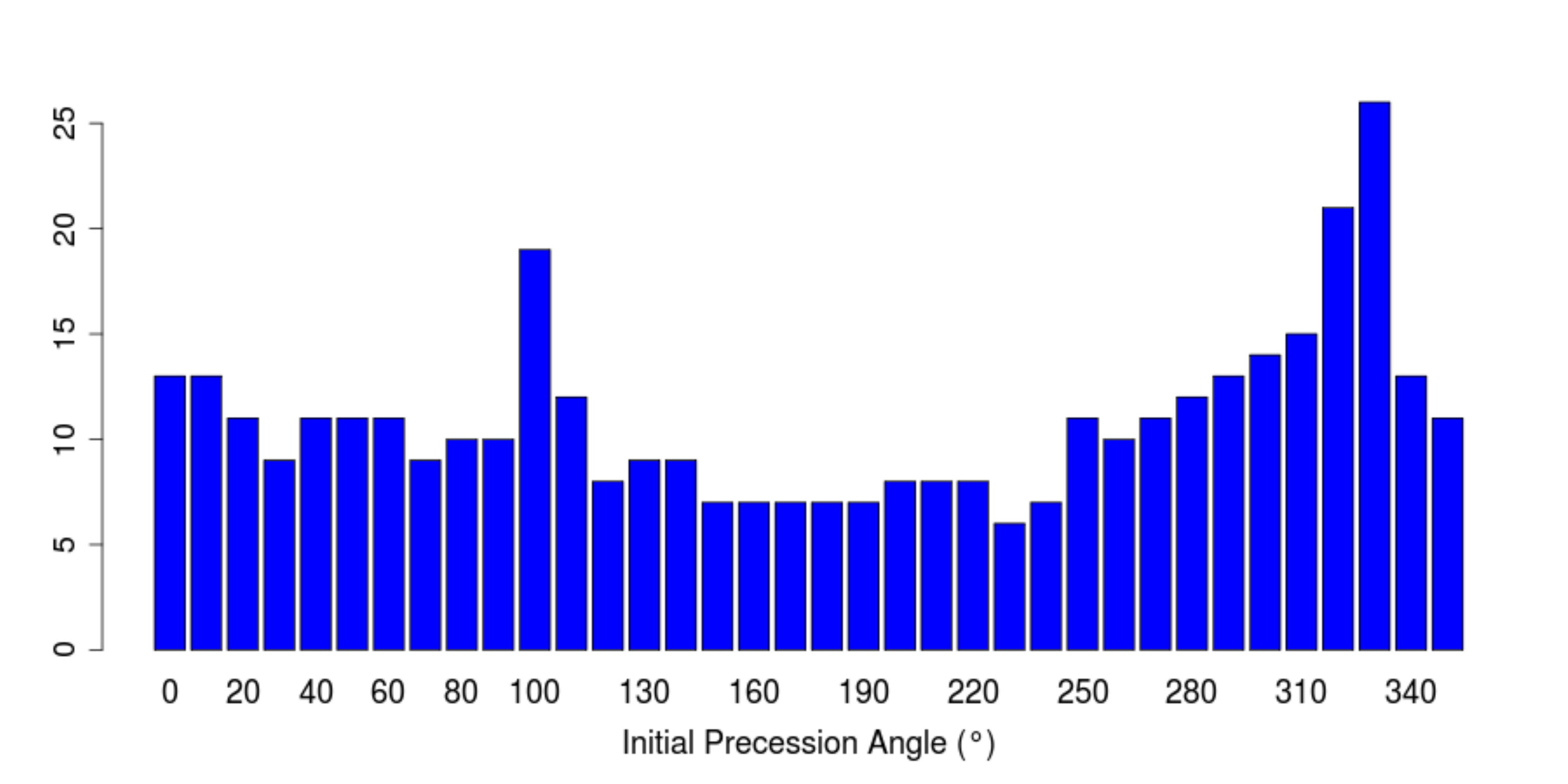}
}
\caption{Number of Gaia observations for the asteroid Apophis with respect to the initial precession angle. Here, we have a great variation of the
number of observations for a single object. Some sets can have more than 25 observations as well as less than 10.}
\label{F:histo_apophis}
\end{figure}

To illustrate the impact of Gaia observations on PHAs orbit, we will consider here the case of the asteroid (99942) Apophis (previously designed
2004 MN$_{\scriptscriptstyle 4}$). This PHA was discovered in June 2004 by R. Tucker, D. Thollen and F. Bernardi at the Kitt Peak observatory in
Arizona. Since the first observations, it was revealed to be a threatening and hazardous asteroid in as much as it reached the level four of Torino
Scale for a possible impact with the Earth in April 2029. Since, new observations ruled out every possibility of collision for this date but this
risk
remains in 2036. The 2029-threat is now just a 2029-close deep encounter within a distance of $\sim 38000$ km with the Earth. Because of this close
encounter, the 2029-post orbit of Apophis is chaotic-like in so far as, the orbit is sensitive to initial conditions, dynamical modelling, etc... Due
to this high sensitivity, some virtual Apophis (clones of the nominal orbit around the nominal value) can be virtual impactors and to quantify and
well appreciate the impact probabilities, it is necessary to well estimate the orbit uncertainties.\\
Apophis has 1366 optical observations and five radar observations spanning 2004-2011 (available at the IAU MPC). Figure \ref{F:histo_apophis} shows
the number of observations that Gaia will provide for this asteroid. One can see that we have inhomogeneous size of sets in as much as we can have
more than 20 observations as well as less than 10 observations. For our simulations, we chose a set with the longest arc length (with 12 Gaia
observations) and with a 5 mas accuracy.
% --------------------------------------------------------------------------------------------------------------
%---------------------------------------------------------------------------------------------------------------

We can first analyse the improvement on the accuracy of the Keplerian elements due to the Gaia contribution. Table \ref{T:stdv} compares the standard
deviation of Apophis'orbital elements with ($\sigma_{\scriptscriptstyle{O+G}}$) and without ($\sigma_{\scriptscriptstyle O}$) Gaia observations. It is
clear that the impact of those space data on Apophis's orbit can be seen through the improvement of the semimajor axis value as the
uncertainty is improved by a factor 1000.\\

%__________________________________________________________________

\begin{table}[h!]
 \begin{center}
  \caption{Stantard deviations of Apophis's keplerian elements without ($\sigma_{\scriptscriptstyle O}$) and with
($\sigma_{\scriptscriptstyle{O+G}}$) 
Gaia observations.}
  \label{T:stdv}
  \begin{tabular}{|c|c|c|}
   \hline
   \hline
            & $\sigma_{\scriptscriptstyle O}$         & $\sigma_{\scriptscriptstyle{O+G}}$ \\
  \hline
   a [A.U.] & $1.9\times 10^{\scriptscriptstyle -08}$ & $6.8\times 10^{\scriptscriptstyle -11}$ \cr
  \hline
   e        & $7.0\times 10^{\scriptscriptstyle -08}$ & $3.9\times 10^{\scriptscriptstyle -09}$ \cr
  \hline
   i [$^\circ$] &  $1.9\times 10^{\scriptscriptstyle -06}$ & $1.2\times 10^{\scriptscriptstyle -07}$ \cr
 \hline
 $\Omega$ [$^\circ$] &  $1.0\times 10^{\scriptscriptstyle -04}$ & $2.2\times 10^{\scriptscriptstyle -06}$ \cr
 \hline
$\omega$ [$^\circ$] &  $1.0\times 10^{\scriptscriptstyle -04}$ & $2.3\times 10^{\scriptscriptstyle -06}$ \cr
 \hline
   M  [$^\circ$] &  $7.4\times 10^{\scriptscriptstyle -05}$ & $6.5\times 10^{\scriptscriptstyle -07}$ \cr

\hline
\hline
  \end{tabular}
 \end{center}
\end{table}

\begin{figure}[htbp!]
\centerline{
	\includegraphics[width=\hsize]{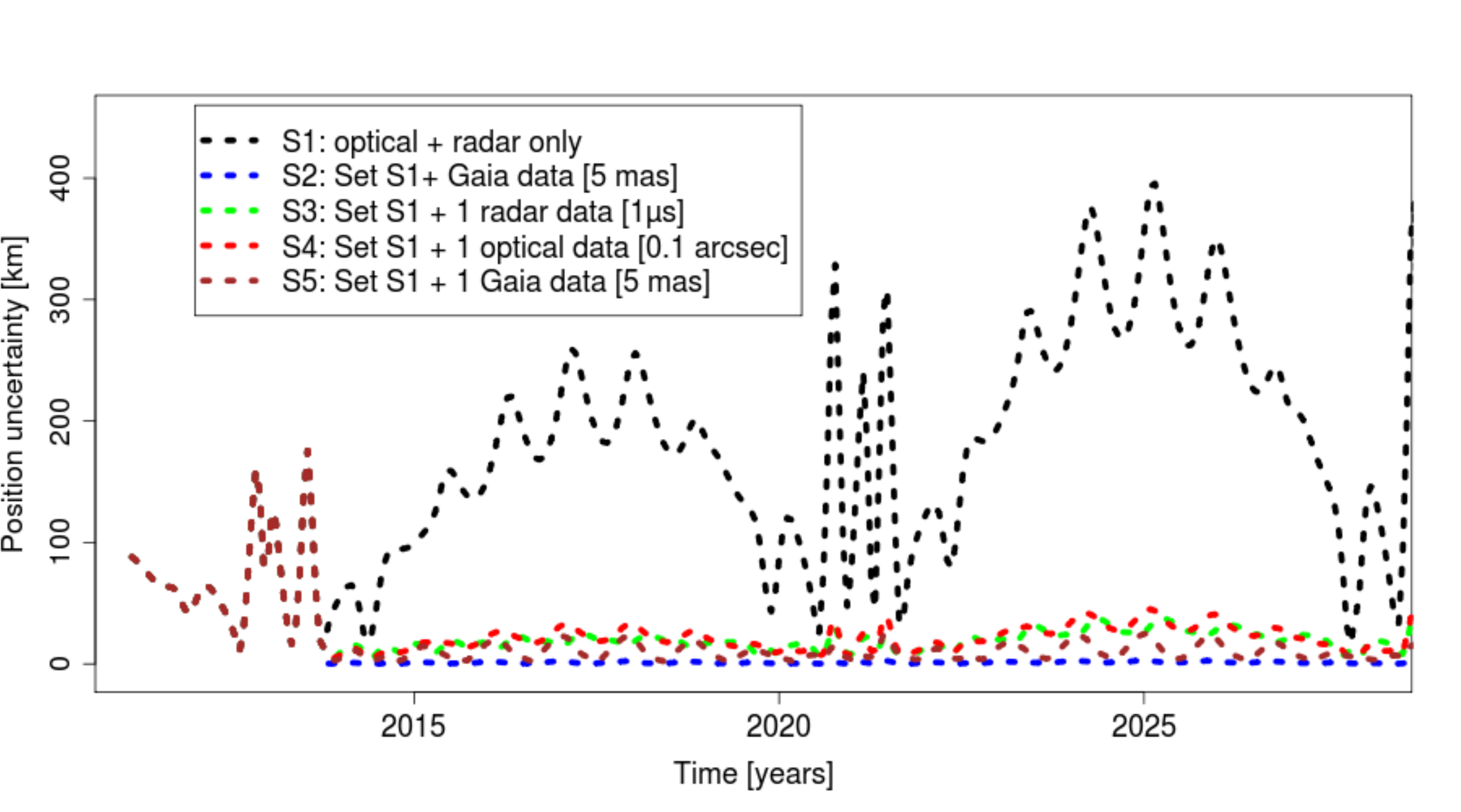}
}
\caption{Evolution of the position uncertainty of asteroid Apophis considering several different sets of observations. While the sets
S$_{\scriptscriptstyle 3}$, S$_{\scriptscriptstyle 4}$ and S$_{\scriptscriptstyle 5}$ lead to the same level of accuracy, the set
S$_{\scriptscriptstyle 2}$ using all Gaia data enable to decrease the position uncertainty down to the kilometer level.}
\label{F:pos_uncertainty}
\end{figure}

The impact of Gaia data can also be analysed through the improvement of the position uncertainty. From a linear propagation of the covariance matrix
(provided by the least square solution), the uncertainty of the keplerian elements is propagated until the date of close approach in 2029. Fig.
\ref{F:pos_uncertainty} shows the comparison of the propagation of nominal orbits obtained from the fit of different sets of observations:
\begin{itemize}
 \item S$_{\scriptscriptstyle 1}$ (\textcolor{black}{-}): using all optical and radar data available;
 \item S$_{\scriptscriptstyle 2}$ (\textcolor{blue}{-}): using set S$_{\scriptscriptstyle 1}$ with additional Gaia data with 5 mas accuracy;
 \item S$_{\scriptscriptstyle 3}$ (\textcolor{green}{-}): using set S$_{\scriptscriptstyle 1}$ with one additional future radar measurement in 2013
with $1\mu $s accuracy (measurement of a timing echo);
 \item S$_{\scriptscriptstyle 4}$ (\textcolor{red}{-}): using set S$_{\scriptscriptstyle 1}$ with one future optical observation done in 2013 with
$0.1$ arcsec accuracy;
 \item S$_{\scriptscriptstyle 5}$ (\textcolor{purple}{-}): using set S$_{\scriptscriptstyle 1}$ and the case that Gaia would provide only one
observation with $5$ mas accuracy.
\end{itemize}

% --------------------------------------------------------------------------------------------------------------
%---------------------------------------------------------------------------------------------------------------

This figure shows that the Gaia data enable to reduce the position uncertainty knowledge down to the kilometer level (set S$_{\scriptscriptstyle 2}$)
and it keeps this value until the close approach. For comparison, the effect of future accurate measurements (radar and optical) can be comparable to
the impact of one future Gaia data. \\

Other simulations can be done to compare the impact of future Gaia data with ground-based measurements by quantifying the position uncertainty at the
date of close encounter. Generally, the uncertainty region is represented in the b-plane or target plane \citep{valsecchi03}. This plane better
represents the state of an asteroid when approching the Earth. It passes through the Earth center and is perpendicular to the geocentric velocity of
the asteroid. Thus, it will have two geocentric coordinates ($\xi,\zeta$). As a consequence, the projection of the ellipsoid uncertainty in this plane
is just an ellipse centered on the nominal value of the geocentric coordinates ($\xi_{\scriptscriptstyle N},\zeta_{\scriptscriptstyle N}$) and with
its semimajor and semiminor axis respectively equal to the standard deviations $3\sigma_{\scriptscriptstyle \zeta}$ and $3\sigma_{\scriptscriptstyle
\xi}$ calculated with a linear propagation of the initial covariance matrix until 2029.\\
Due to this close approach, the orbit of Apophis will be altered and both Apophis and the Earth are expected to be in the same position after some
revolutions of Apophis around the Sun and many years later. The most famous resonant return occurs in 2036 where after 6 revolutions of Apophis and 7
years later, both objects will meet again. As the 2029-post orbit of Apophis is chaotic, some clones of Apophis (simulating by Monte-Carlo the
present orbital uncertainty) can lead to impact with the Earth at some resonant return and the pre-images of those impacts in the b-plane are called
keyholes \citep{chodas99}. The most famous keyhole is the 2036-keyhole with a size around $600$ m. They can be primary keyholes if they are spawned by
one close approach and secondary if they are spawned by two consecutive close encounters. So, the risk can be estimated by comparing the keyhole
position with the size of the ellipse uncertainty in the b-plane. A better knowledge of the region uncertainty is necessary to prepare some deflection
missions in case there is an important collision threat.\\
The size of the region uncertainty, in the ($\xi,\zeta$) plane, will depend on the kind of measurements available. Table \ref{T:b_plane_uncertainty}
presents the size of the ellipse uncertainty using the different sets S$_{\scriptscriptstyle i}$ of observations as explained above. Even if Gaia
would provide only one observation, the gain in accuracy would be unprecedented by comparison with the gain obtained with optical or radar data.
While the impact of one Gaia data can be compared to the effect of one radar measurement, one set of Gaia observations can bring the
uncertainties around the kilometer level.

\begin{table}[h!]
 \begin{center}
  \caption{Uncertainties ($\sigma_{\scriptscriptstyle \xi}$,$\sigma_{\scriptscriptstyle \zeta}$) on the 2029 b-plane of Apophis considering various
sets
S$_{\scriptscriptstyle i}$ of observations. }
  \label{T:b_plane_uncertainty}
  \begin{tabular}{|c|c|c|c|c|c|}
   \hline
   \hline
            & S$_{\scriptscriptstyle 1}$ & S$_{\scriptscriptstyle 2}$ & S$_{\scriptscriptstyle 3}$ & S$_{\scriptscriptstyle 4}$ &
S$_{\scriptscriptstyle 5}$    \cr
  \hline
   $\sigma_{\scriptscriptstyle \xi}$ (km)& $10$  & $0.3$ & $7$ & $8$ & $6$\cr
  \hline
   $\sigma_{\scriptscriptstyle \zeta}$ (km) & $240$ & $1.6$ & $10.5$ & $24$ & $11.5$\cr   
  
\hline
\hline
  \end{tabular}
 \end{center}
\end{table}

\begin{figure}[htbp!]
\centerline{
	\includegraphics[width=\hsize]{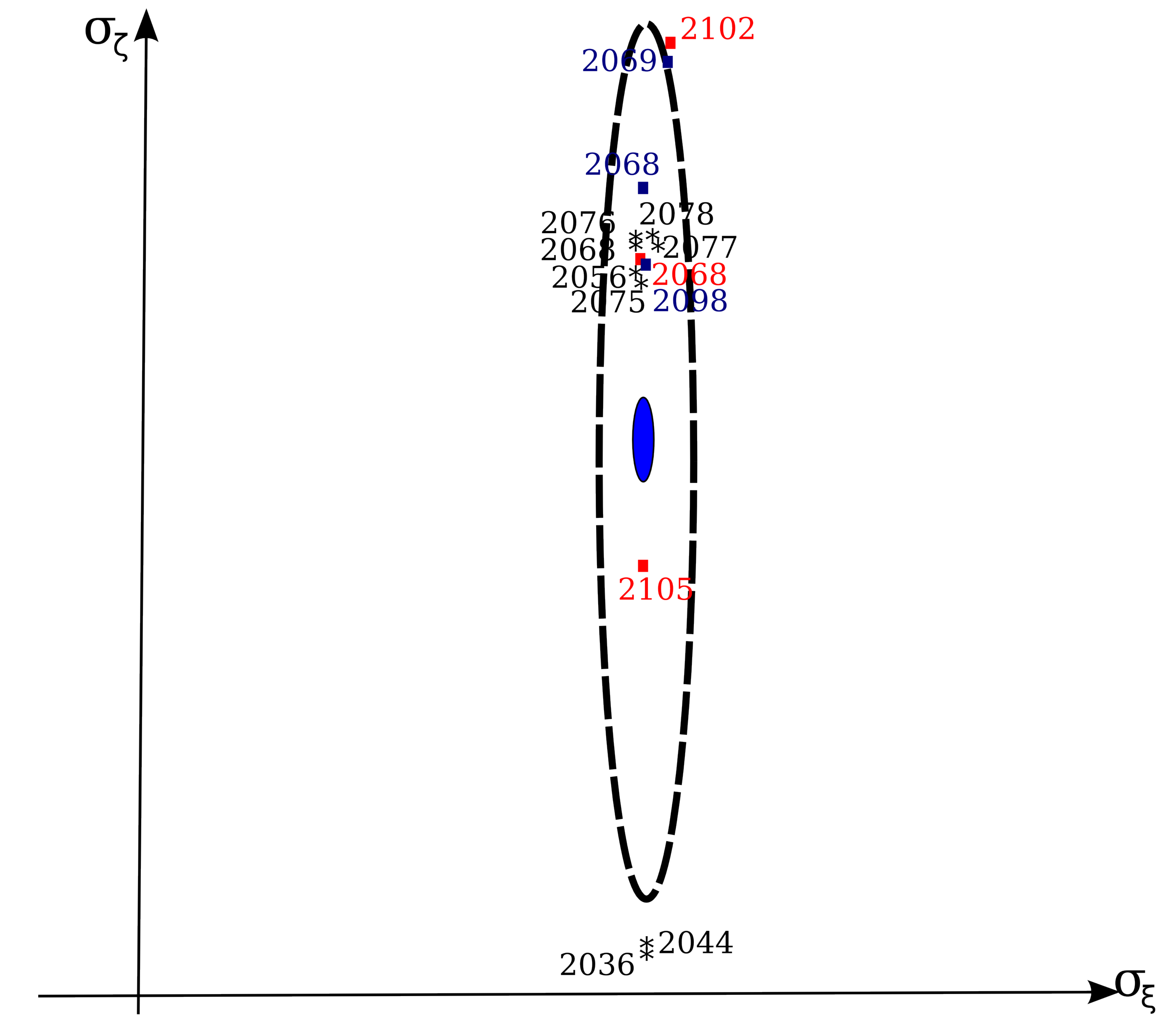}
}
\caption{$3\,\sigma$ ellipse uncertainty on the 2029 b-plane of Apophis and position of the center of primary (\textcolor{black}{$\star$}) and
secondary keyholes leading to collision at ascending node (\textcolor{red}{$\blacksquare$}) and descending node 
(\textcolor{blue}{$\blacksquare$}). The dotted ellipse is computed using set S$_{\scriptscriptstyle 1}$ and the filled one using
set S$_{\scriptscriptstyle 2}$. The coordinates are expressed in $\sigma$ units.}
\label{F:b_plane}
\end{figure}

Finally, it could be interesting to map the primary and secondary keyholes in the 2029 b-plane in order to illustrate the improvement due to the Gaia
data. Figure \ref{F:b_plane} shows the position of the center of those keyholes. Those positions were numerically computed using Monte-Carlo
technique and using the Lie integrator \citep{bancelin11}. So, as the region uncertainty shrinks thanks to the Gaia data (small ellipse), the
collision probabilites will also decrease as the distance between the keyholes center and the center of the ellipse increases. 

% --------------------------------------------------------------------------------------------------------------
%---------------------------------------------------------------------------------------------------------------

%__________________________________________________________________

%--------------------------------------------------------------------------------------------
\section{Gaia-FUN-SSO network}
\label{S:network}

%  \textit{Besoins obs sol (WT)\\
%     statut du reseau (WT)\\
%     composition (WT)\\
%  Statistiques sur les alertes NEAs (DB) }

During the mission, various unidentified objects will be observed by the satellite. Because of the scanning law, at the epoch of these discoveries,
those objects will have at least two observations separated by approximately $\Delta\,t\,\sim\,1.5$ hours. But, as Gaia is not a follow-up satellite,
the newly discovered asteroids can be rapidly lost if there is no follow-up from the Earth. Among the potential alerts, we expect some NEOs (and
amoung them PHAs) to be discovered. We also expect the discovery of  several Inner-Earth Objects (IEOs) due to the L2 positionning of the probe and
of the 45 degrees solar elongation which will allow it to investigate inside the Earth orbit. We can also  expect the discovery of  new comets.
In order to be ready to handle those alerts, we first have to statistically quantify the number of unknown NEOs that could be discovered by Gaia. 
In a first approach, using a synthetic population of NEOs \citep{bottke02}, we do expect a small number of alerts ($\sim 1$ alert every $4$ days) by
comparing the number of known and synthetic NEOs that would be observed by the satellite during the 5 years mission (Fig. \ref{F:all_nea}).

% --------------------------------------------------------------------------------------------------------------
\begin{figure}[htbp!]
\centerline{
	\includegraphics[width=\hsize]{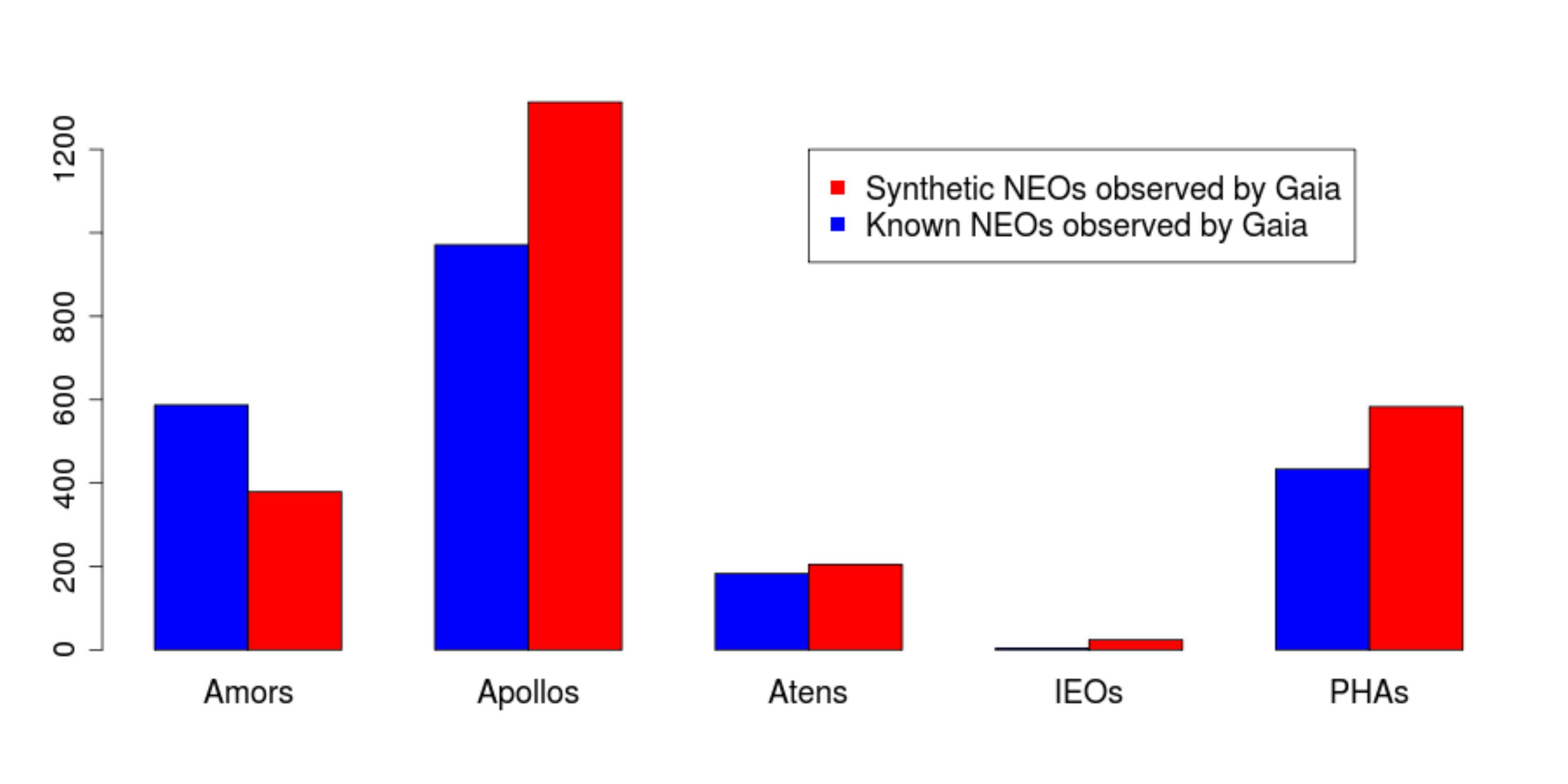}
}
\caption{Number of known and synthetic NEOs that would be observed by the satellite.}
\label{F:all_nea}
\end{figure}
%---------------------------------------------------------------------------------------------------------------

According to the previous considerations upon the interest of a ground-based follow-up network, we have set up a ground-based network of observing
sites labelled Gaia-FUN-SSO (standing for Gaia Follow-Up Network for Solar System Objects).  This network included nineteen locations at the beginning
of 2011 but several more stations are still expected in order to have a large geographical coverage (candidate sites can get in touch with us at the
address gaia-fun-sso@imcce.fr). The telescope diameters of the network are spanning from 0.25 to 2.4 m; four telescopes have large field, which will
be useful for recoveries, and five are robotics ones, which will be precious for observations on alert. 
Since the goal is mainly to perform astrometric measurements, the standard specificities of telescopes are expected to be a field of view of at least 10 arcmin, pixel size at less than 1 arcsec, and limiting magnitude around 20. But, since we certainly need to search for new discovered objects in quite large field and larger field even with bigger pixel size will be very useful. 

The role of this network will be to improve the orbit of some objects and to enable Gaia to identify them during a further scan. This network is
structured around a central node which will convert raw Gaia data into ephemerides useful for observations and will collect the data. All the
measurements performed by this network will be sent to the Minor Planet Center and will thus allow the update of the database of auxiliary data used
in the Gaia system  to perform the identification of SSO. A first workshop has been held in Paris in November 2010 and resulted in several discussions
among the member of the network; proceedings are accessible at the address:  gaia-fun-sso.imcce.fr. 

% --------------------------------------------------------------------------------------------------------------
\begin{figure}[htbp!]
\centerline{
	\includegraphics[width=\hsize]{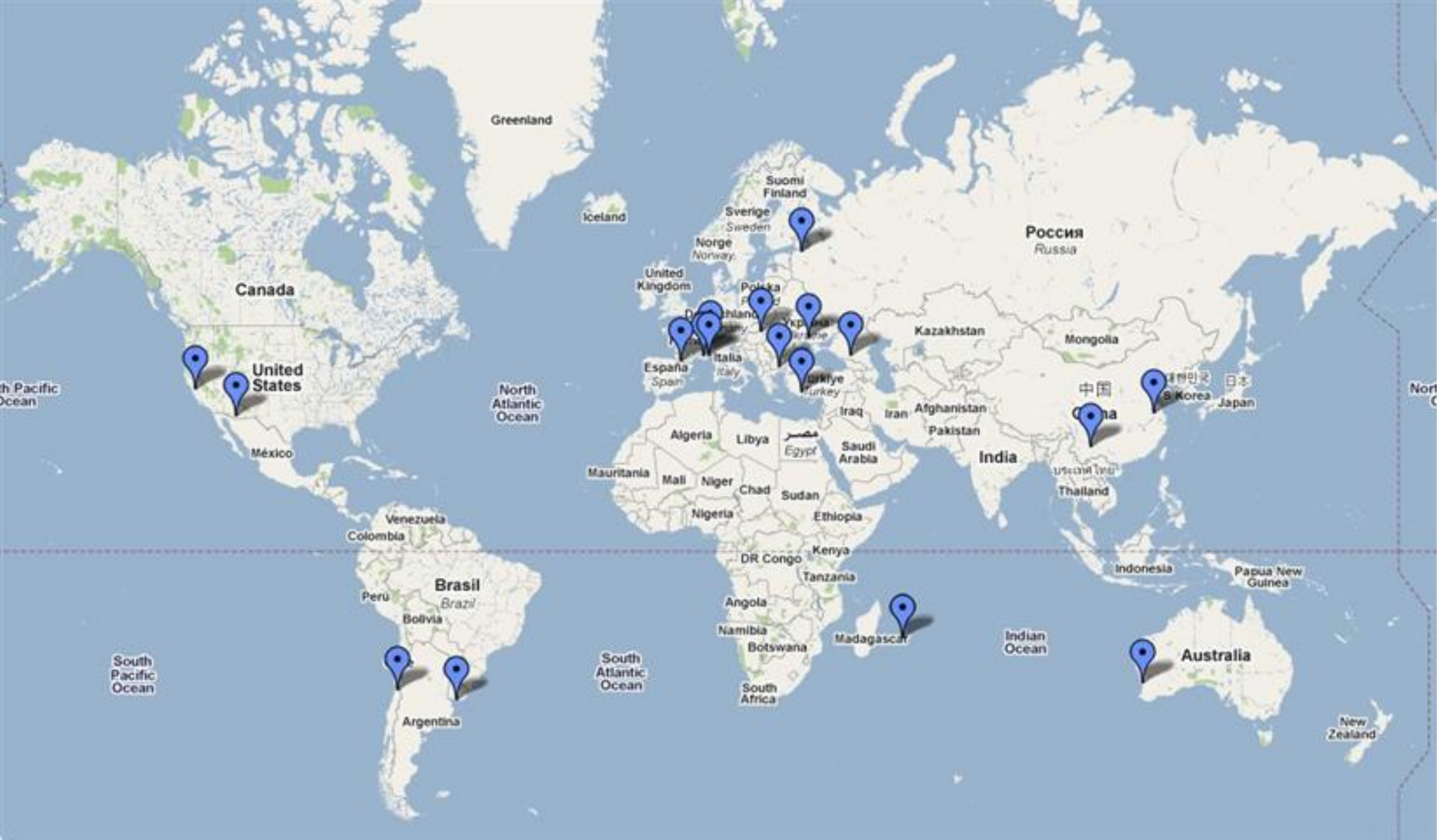}
}
\caption{Observing sites of the Gaia-FUN-SSO network in  May 2011}
\label{F:all_nea}
\end{figure}
%---------------------------------------------------------------------------------------------------------------

%--------------------------------------------------------------------------------------------

\section{Synergy ground/space data}
\label{S:synergy}

% DB+WT
% 
%  \textit{    combinaisons obs sol+espace (radar, extension de la periode d'obs)\\
%     effet parallaxe\\
%     statist ranging\\
% }

\begin{figure}[htbp!]
\centerline{
	\includegraphics[width=\hsize]{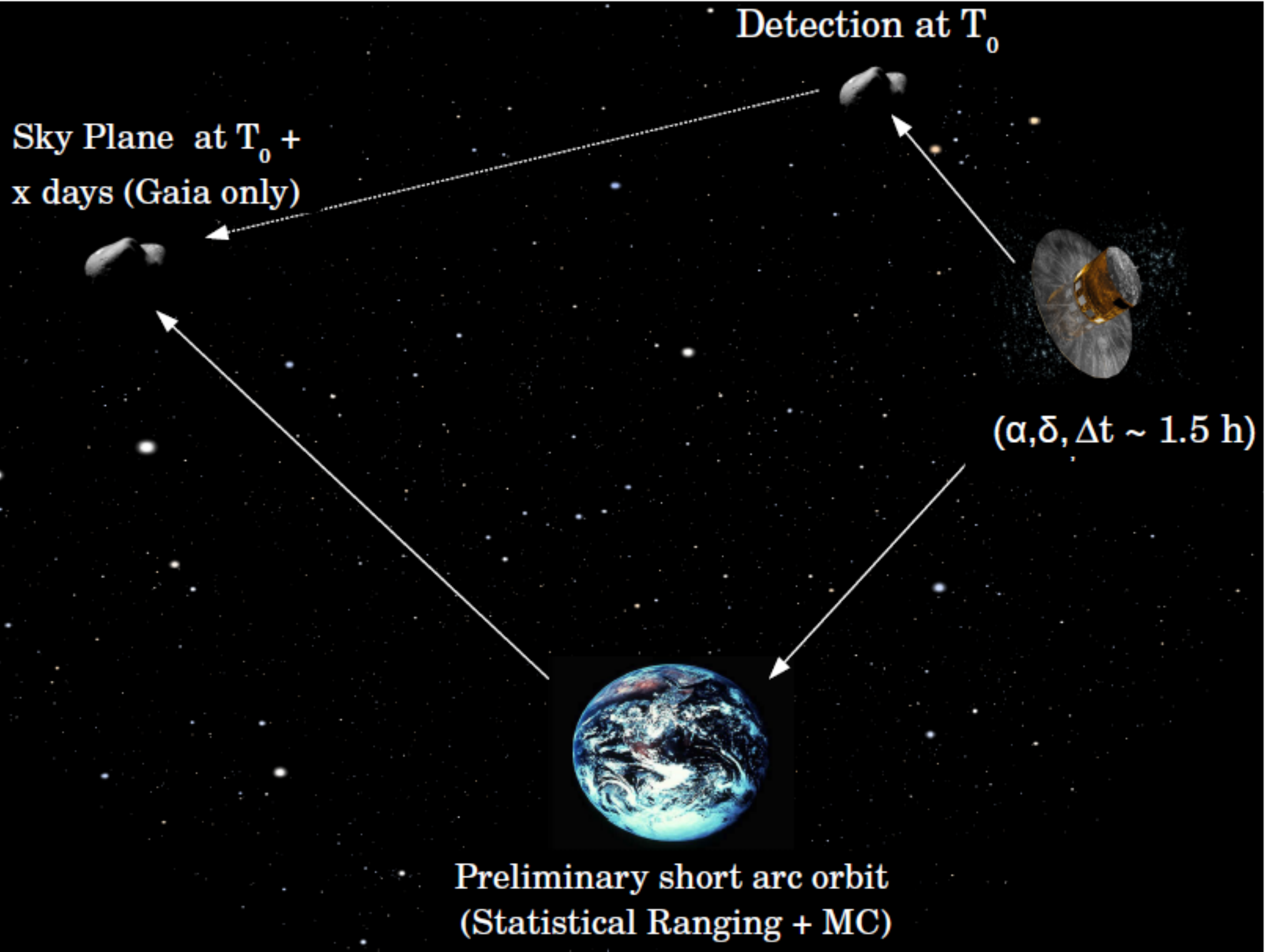}
}
\caption{Strategy of recovery from Earth for newly discovered PHAs. Gaia will provide two observations before sending the coordinates to Earth
within
24 hours where a short preliminary arc orbit, compatible with the observations, can be computed using the Statistical Ranging method. 
(MC denotes Monte-Carlo Technique). Thus, an ($\alpha$, $\delta$) distribution can be computed few days after the discovery of the asteroid by
Gaia.}
\label{F:strategy}
\end{figure}

When an alert occurs, we have to know where to look in the celestial sphere and how much time we have in order to recover, from the Earth, an
unidentified asteroid observed by Gaia. Knowing the threat of PHAs, we can not afford to lose them if no strategy is established. A way to deal with
potential alerts can be represented in Fig. \ref{F:strategy}: If an unidentified PHA is observed by Gaia, the satellite can send an alert to the Earth
within 24 hours. Then, a short preliminary arc orbit, compatible with the Gaia observations, can be computed using the Statistical Ranging method
\citep{virtanen01, muinonen_pisa}. This method is based on estimating the gaiacentric distance using Monte-Carlo technique with at least two
observations. It will provide the orbital elements compatible with the Gaia observations and propagate each orbit to a given date. Then, from the
($\alpha$,$\delta$) distribution computed few days after its discovery, we can extract the maximum likelyhood of this distribution. We can then just
center a telescope field of view on this maximum likelyhood so that observers can be able to know which part of the sky to scan and how much time they
have until the asteroid is lost.\\

% --------------------------------------------------------------------------------------------------------------
%---------------------------------------------------------------------------------------------------------------

\begin{figure}[htbp!]
\centerline{
	\includegraphics[width=\hsize]{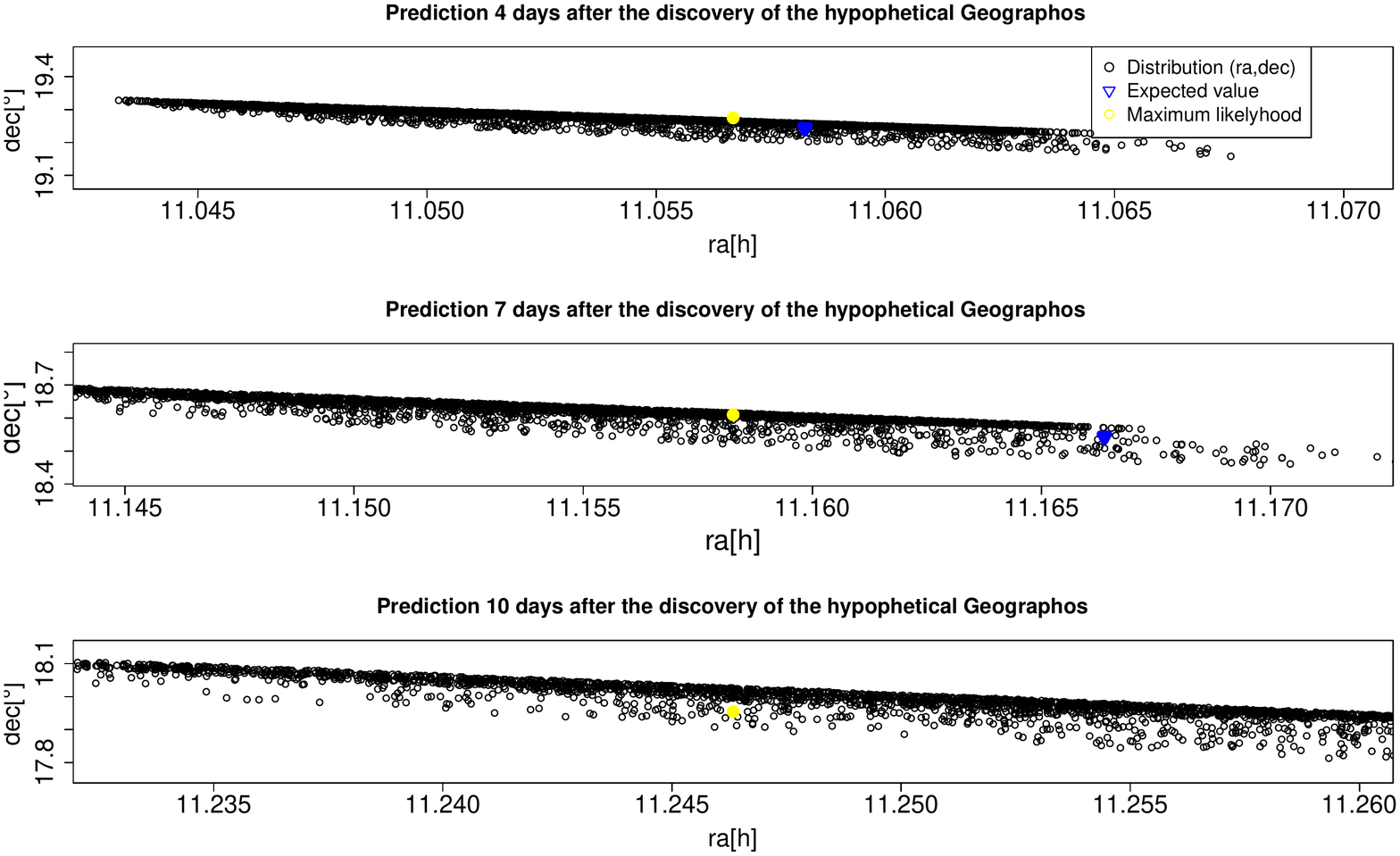}}
\caption{Prediction on the sky plane of a hypothetical Geographos discovered by Gaia, until 10 days after its discovery by Gaia.}
\label{F:distrib_sans_geo}
\end{figure}

As an example, we considered an hypophetical PHA, Geographos, that would be discovered by Gaia. Figure \ref{F:distrib_sans_geo} shows the
($\alpha$,$\delta$) distribution on the sky plane (\textcolor{black}{$\circ$}) until 10 days after the discovery of Geographos. Each window is
centered on the maximum likelyhood (\textcolor{yellow}{$\bullet$}) and the size of the window is the size of a $24\times 24$ arcmin telescope field of
view. So, the asteroid can still be recovered until 7 days after its discovery because the true value (\textcolor{blue}{$\blacktriangle$}),
computed from the real initial state of Geographos, lies in this window for this given field of view.
% --------------------------------------------------------------------------------------------------------------
%---------------------------------------------------------------------------------------------------------------

\begin{figure}[htbp!]
\centerline{
	\includegraphics[width=\hsize]{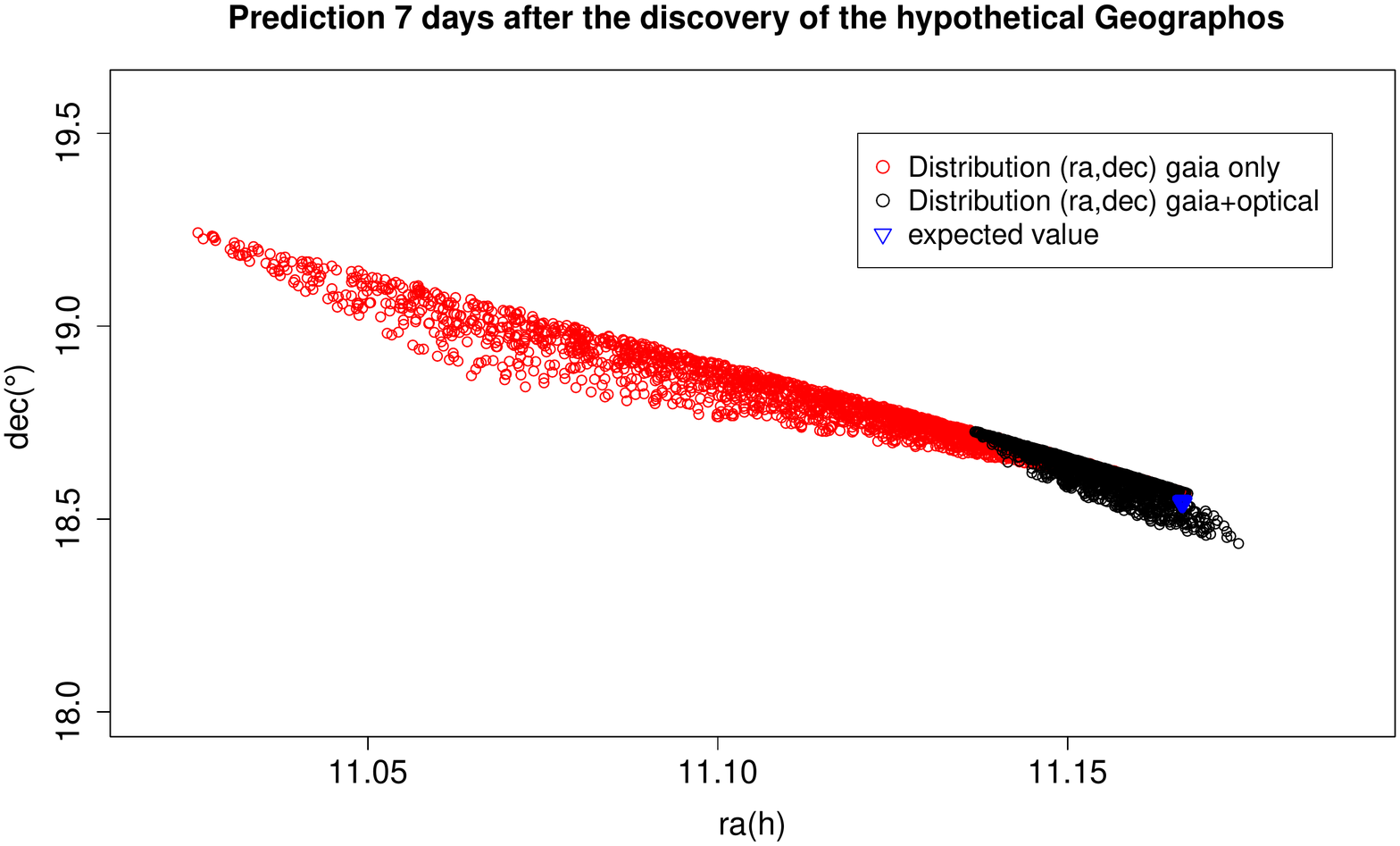}
}
\caption{Distribution ($\alpha$,$\delta$) considering additional ground-based data, two days after the discovery of the hypothetical Geographos by
Gaia.}
\label{F:distrib_avec_geo}
\end{figure}

Finally, when the asteroid is recovered from Earth, it will be followed, at least, during one night. Thus, optical data can be done and can be
combined with the space data in order to improve the ($\alpha$,$\delta$) prediction in the sky plane. We considered four observations made during that
night with a $0.5$ arcsec accuracy, two days after its discovery by the satellite. The optical data enable to better constrain the preliminary short
arc orbit and as seen in Fig.\ref{F:distrib_avec_geo}, the parallax effect allows a better ($\alpha$,$\delta$) prediction as the size of the
distribution is well-reduced (light circles), compared to the distribution obtained only with Gaia data (black circles).
% --------------------------------------------------------------------------------------------------------------
%---------------------------------------------------------------------------------------------------------------

%-------------------------------------------------------------------------------------------------------
\section{Conclusion}
\label{conclusion}
%%DB+WT(+DH)

We have given a broad overview of results and actions connected to the astrometry of asteroids and NEOs with Gaia. This includes the orbit
improvement, mass determination, test of GR. This paper also presented the usefulness of Gaia data thanks to an unprecedented data accuracy reached.
Orbit of NEOs and PHAs could really be improved, even if the number of observations provided by the satellite is faint. This improvement can be shown
through the improvement of orbital elements, position uncertainty and even for close-approach statistics.\\
\indent Even if Gaia won't be a big NEOs discoverer and is not a follow-up mission, a strategy has to be settled in order to be able to recover newly
discovered PHAs from Earth. Statistical tools can enable observers to know where to focus on the celestial sphere with only two Gaia data. Besides,
the parallax effect, with addionnal ground-based data, will allow a better follow-up from Earth. 

\bigskip
\noindent{\em Acknowledgements:} {The authors wish to thank J. Blanchot and M. Sylvestre -- master trainees at IMCCE -- for her work on GIBIS detection, C. Ordenovic and F. Mignard (OCA) for providing the CU4 Solar System Simulator, and all the colleagues from Gaia DPAC CU4/SSO and REMAT groups at large for fruitful discussions. }
%-------------------------------------------------------------------------------------------------------



%% References

\bibliographystyle{model2-names}
\bibliography{Gaia_pisa}






% \end{thebibliography}

\end{document}